\documentclass{article}
\usepackage{spconf,amsmath,graphicx}

\copyrightnotice{978-1-6654-7189-3/22/\$31.00~\copyright2023 IEEE}

\usepackage[table]{xcolor}
\usepackage{booktabs}
\usepackage[hidelinks]{hyperref}
\usepackage{multirow}

\definecolor{Gray}{gray}{0.9}

\title{How Does Pre-trained Wav2Vec 2.0 Perform on Domain-Shifted~ASR?\\ An Extensive Benchmark on Air Traffic Control Communications}
\name{
\begin{tabular}{c}
Juan Zuluaga-Gomez$^{~\star, \dagger, \ddagger}$, 
Amrutha Prasad$^{~\dagger, \mathparagraph}$, 
Iuliia Nigmatulina$^{~\dagger}$, 
Seyyed Saeed Sarfjoo$^{\dagger}$, \\
Petr Motlicek$^{\,\dagger, \mathparagraph}$, 
Matthias Kleinert$^{\,\mathsection}$, 
Hartmut Helmke$^{\,\mathsection}$, 
Oliver Ohneiser$^{\,\mathsection}$, 
Qingran Zhan$^{\|}$
\end{tabular}
\thanks{Copyright 2023 IEEE. Published in the 2022 IEEE Spoken Language Technology Workshop (SLT) (SLT 2022), scheduled for 19-22 January 2023 in Doha, Qatar. Personal use of this material is permitted. However, permission to reprint/republish this material for advertising or promotional purposes or for creating new collective works for resale or redistribution to servers or lists, or to reuse any copyrighted component of this work in other works, must be obtained from the IEEE.}
\thanks{$^{\star}$Corresponding author: juan-pablo.zuluaga@idiap.ch}
}

\address{
  $^{\dagger}$ Idiap Research Institute, Martigny, Switzerland \\
  $^{\ddagger}$ Ecole Polytechnique Federale de Lausanne (EPFL), Switzerland \\
  $^{\mathparagraph}$ Brno University of Technology, Brno, Czech Republic \\
  $^{\mathsection}$ German Aerospace Center (DLR), Institute of Flight Guidance, Braunschweig, Germany \\
  $^\|$School of Information and Electronics, Beijing Institute of Technology, Beijing, China \\
}

\begin{document}
\ninept
\maketitle
\begin{abstract}
Recent work on self-supervised pre-training focus on leveraging large-scale unlabeled speech data to build robust end-to-end (E2E) acoustic models (AM) that can be later fine-tuned on downstream tasks e.g., automatic speech recognition (ASR). Yet, few works investigated the impact on performance when the data properties substantially differ between the pre-training and fine-tuning phases, termed domain shift. We target this scenario by analyzing the robustness of Wav2Vec 2.0 and XLS-R models on downstream ASR for a completely unseen domain, air traffic control (ATC) communications. We benchmark these two models on several open-source and challenging ATC databases with signal-to-noise ratio between 5 to 20\,dB. Relative word error rate (WER) reductions between 20\% to 40\% are obtained in comparison to hybrid-based ASR baselines by only fine-tuning E2E acoustic models with a smaller fraction of labeled data. We analyze WERs on the low-resource scenario and gender bias carried by one ATC dataset.
\end{abstract}
\begin{keywords}
Automatic speech recognition, Wav2Vec 2.0, self-supervised pre-training, air traffic control communications.
\end{keywords}
\section{Introduction}
\label{sec:intro}

A lot of recent work on end-to-end (E2E) acoustic modeling including automatic speech recognition (ASR) exploits self-supervised learning (SSL) of speech representations~\cite{schneider2019wav2vec} including autoregressive models~\cite{oord2018representation, baevski2019vq} and bidirectional models~\cite{baevski2020wav2vec,chen2021wavlm}. Self-supervised learning is a training technique capable of leveraging large-scale unlabeled speech to develop robust acoustic models~\cite{baevski2020wav2vec,baevski2022data2vec}. In fact,~\cite{baevski2021unsupervised} explores a way to perform ASR without any labeled data in a complete unsupervised fashion. In a standard setup, E2E models trained by SSL are later fine-tuned on downstream tasks with much fewer labeled samples compared to standard supervised learning. By applying SSL, these systems have dramatically improved ASR performances on English speech datasets~\cite{baevski2020wav2vec}, such as LibriSpeech~\cite{panayotov_librispeech_ICASSP2015}. Similarly, performance on cross-lingual speech recognition is largely improved by SSL~\cite{zhang2021xlst,babu2021xls}. It can be inferred that SSL-based pre-training allows models to capture a good representation of acoustics, which can be leveraged across different languages for ASR. 
\vspace{0.1cm}

This work reviews the robustness of two well-known E2E acoustic models trained by SSL (i.e., Wav2Vec 2.0 and XLS-R) on a completely unseen domain: air traffic control (ATC) communications. ATC deals with aircraft guidance in the air and on the ground via voice communications between air traffic controllers (ATCOs) and pilots. 
These communications are well-defined by grammar and vocabulary~\cite{allclear} that must be followed to provide a safe and reliable flow of air traffic while keeping operating costs as low as possible.
Despite the interest of ASR for ATC, there is not a fully functional ASR engine on the market due to: (i)~lack of performance, i.e., under 5\% WER is required,\footnote{There is a clear threshold between enhancing ATCOs' productivity and delaying them in their tasks due to poor ASR, see~\cite{ohneiser2021robust}.} and (ii)~lack of large-scale annotated speech (less than 50\,hrs of open-source databases) and its high production cost makes it almost impractical~\cite{zuluagagomez21_interspeech}.

\subsection{Contribution and motivation}

Only a few previous works intended to measure the effect of domain mismatch between pre-training and fine-tuning phases of E2E models~\cite{hsu2021robust}. However, we can still categorize all databases as either read, spontaneous or conversational speech. On the contrary, ATC speech does not fit in any of these three categories due to its uniqueness, e.g., ruled by a very well-defined grammar. Our contributions\footnote{Our code is stored in the following public GitHub repository: \url{https://github.com/idiap/w2v2-air-traffic}} cover the domain mismatch scenario by answering the three questions below. 
\vspace{0.2cm}

\noindent \textbf{(i)~How robust pre-trained E2E models are on new domains like ATC?} Our results (see Table~\ref{tab:results}) ratified that E2E models pre-trained by SSL (e.g., Wav2Vec 2.0) learn a strong representation of speech. Fine-tuning on a downstream task (e.g., ASR) is computationally less expensive than flat-start training, and it requires less in-domain data to achieve on par WERs compared to traditional hybrid-based ASR systems. 
We also perform experiments with multilingual E2E models, i.e., XLS-R~\cite{babu2021xls}. 
We hypothesize that pre-trained multilingual models perform better on ATC speech data that contains accented English (i.e., LiveATC-Test and ATCO2-Test sets). Potentially, due to the strong speech representation acquired during SSL phase, which translates into a more accent-agnostic AM.
\vspace{0.2cm}

\noindent \textbf{(ii)~How much in-domain ATC labeled data is needed to fine-tune an E2E model that reaches on-par performance with regard to hybrid-based models?} We perform a comparative study ranging from 5\,minutes (few-shot learning) to $\sim$15\,hrs of labeled speech (i.e., from 100 to 15k utterances). 
In addition, we investigate the impact of integrating an in-domain language model with beam search rather than using simple greedy decoding.
With the aim of open science and fostering research in ATC, we also introduce baselines\footnote{Our code is stored in the following public GitHub repository: \url{https://github.com/idiap/w2v2-air-traffic}} on two well-known public ATC corpora.
\vspace{0.2cm}

\noindent \textbf{(ii)~How robust are E2E models on speakers with different gender?} E2E models such as Wav2Vec 2.0 have experienced exponential interest in the research community. However, little investigation has been carried out about estimating the WERs gap produced by gender disparities. To name a few~\cite{meng2022don,zanonboito22_interspeech,riviere2021asr4real}. In this work, we perform an analysis by fine-tuning E2E models with ATC audio from different genders. We study this on a free and public ATC corpus where gender labels are provided.
\vspace{0.2cm}

We believe this work is impactful because the ASR field is advancing in an outpacing manner, where each month many large-scale speech models (LSSM) are poured into the research field with outstanding performances in well-known corpora, e.g., LibriSpeech~\cite{panayotov_librispeech_ICASSP2015}. Nonetheless, little has been examined in many other domains, such as ATC communications. Thus, it is of particular interest to evaluate and assess the performance of these LSSM on \emph{`lagged'} fields.



\section{Related Work}
\label{sec:related-work}

Robust ASR stands as a promising tool for aiding ATCOs and ATC, in several ways.
For example, reducing ATCOs' workload~\cite{helmke2016reducing} by automatizing several of their daily tasks.\footnote{One example is to detect named entities from voice communications. These entities are later parsed into ATCOs workstations. Using ASR reduces the overall latency of this simple, yet important procedure.} Another by-product of introducing ASR tools in ATC is the increase in airspace safeness and reduction of environmental impact caused by ATC operations. This is why the European Union (EU) has been funding different projects intended to bring closer speech and text-based technologies to ATC. \mbox{MALORCA} project concluded that ATCOs's workload can be reduced significantly by integrating ASR systems, while boosting their efficiency~\cite{helmke2017increasing}. 
ATCO2\footnote{\url{https://www.atco2.org/}}is also a well-known project that developed a pipeline~\cite{kocour2021automatic} for automatic collection and pre-processing of large quantities of ATC audio data. Their main focus covered downstream task such as ASR~\cite{zuluagagomez20_interspeech}, named-entity recognition~\cite{zuluaga2020automatic}, and acoustic and text-based speaker role recognition~\cite{prasad2021grammar, zuluaga2022bertraffic}. HAAWAII\footnote{\url{https://www.haawaii.de}} project develops a reliable and adaptable solution to automatically transcribe voice utterances issued by both ATCOs and pilots. Still, all previous research only investigates either standard supervised or semi-supervised~\cite{imseng2014exploiting,srinivasamurthy2017semi, kleinert2018semi} hybrid-based ASR systems.
\vspace{0.1cm}

Information uttered in ATC communications contain a diverse set of special entities, also called named entities. Some examples are callsigns, values and units. The most important and critical is the callsign, composed of an \textit{airline designator}, a set of numbers and letters, e.g., \verb|TVS12AB| spelled as \textit{``\textcolor{blue}{SKYTRAVEL ONE TWO ALFA BRAVO}}. The correct recognition of such key entities is crucial, as further it is used to extract target information from the conversations to assist ATCOs. Thus, it is almost mandatory for ASR engines to provide considerable low WERs. Additionally, the communications are mostly carried over noisy audio channels, usually below 15\,dB SNR, which is the default in ATC environments. Taking into account these considerations, the \textit{`ideal ASR engine'} should aim at preventing error propagation to the fullest extent, which can turns into misleading information passed to sub-systems at the next stages. 
\vspace{0.1cm}

We redirect the reader to a general overview on spoken instruction understanding for ATC in~\cite{lin2021spoken} and latest work on hybrid-based ASR for ATC in~\cite{kocour2021automatic,zuluaga2020automatic}. In~\cite{zuluagagomez21_interspeech} contextual information (also known as contextual biasing) via n-grams composition (in the HCLG graph\footnote{In hybrid-based ASR, the different knowledge sources are represented via weighted finite-state transducers (WFST) and then merged in a final decoding graph i.e., \textit{HCLG}. H, C, L and G are the hidden Markov models, context dependency, lexicon, and LM or grammar, correspondingly.}) is merged with semi-supervised learning techniques to further decrease word error rates (WER) on an ASR designed for ATC. Boosting of contextual knowledge during and after decoding has also been explored in~\cite{motlicek2010english,kocour21_interspeech, nigmatulina2021improving,nigmatulina2022two}, where a set of target n-grams are added to further decrease WERs.
\vspace{0.1cm}

Despite the recent success of mixing SSL acoustic pre-training on E2E architectures for ASR, there has not been a comparative study between traditional hybrid-based and E2E acoustic modeling targeted to ATC. First, hybrid-based ASR modeling is based on a disjoint optimization of separate models i.e., AM, LM and a lexicon (e.g., phoneme-based). State-of-the-art (SOTA) models are trained with lattice-free maximum mutual information (LF-MMI) loss~\cite{povey2016purely} which relies on alignments produced by a previously trained HMM-GMM model~\cite{povey2016purely}. Second, E2E systems model AM and LM jointly, and they are mostly trained with connectionist temporal classification (CTC) loss~\cite{graves2006connectionist} (enabling alignment-free training). In~\cite{vyas2021lattice}, it is compared CTC and LF-MMI adaptation of pre-trained models. Recently, attention-based (e.g., Transformers) have become the \textit{de facto} choice for AM~\cite{baevski2020wav2vec,babu2021xls,baevski2020effectiveness}. However, only few studies focused on domain shift during pre-training and fine-tuning or the impact of noisy speech on AM~\cite{zhu2022noise}. For instance,~\cite{hsu2021robust, kawakami2020learning, vyas2021comparing} perform experiments similar to ours, addressing the domain-shift scenario between pre-training and fine-tuning phases. Yet, these databases still fall into read, spontaneous or conversational speech.


\section{Datasets and Experimental Setup}
\label{sec:datasets}

This research experiments with seven datasets in the English language with various accents, speech rate, and data quality. With the aim of encouraging open research on ATC,\footnote{General research on ASR for ATC has lagged behind due to privacy clauses and contracts. Ongoing and former projects prohibit code release.} we experimented with four public databases,\footnote{See the GitHub repository for further information and baselines.} as referenced in Table~\ref{tab:datasets}. To the author's knowledge, this is the first work that open sources code in the field of robust ASR targeted to ATC. 

\subsection{Private databases}

\qquad \textbf{NATS and ISAVIA}: the audio data is collected and annotated by air navigation service providers (ANSPs) for HAAWAII project. The two datasets are, (i)~London approach (NATS) and (ii)~Icelandic en-route (ISAVIA). In total, there are 32\,hrs of manually transcribed data for training and 2\,hrs for testing. Both datasets are cataloged as good quality speech sampled at 8\,kHz. Further details in Table~\ref{tab:datasets}. 
\vspace{0.1cm}

\textbf{LiveATC-Test:} the test set is gathered from LiveATC\footnote{Streaming audio platform that gathers VHF ATC communications.} data recorded from publicly accessible VHF radio channels, as a part of ATCO2 project~\cite{zuluagagomez21_interspeech, nigmatulina2022two}, and includes pilot and ATCO recordings with accented English from airports located in U.S., Czech Republic, Ireland, Netherlands, and Switzerland. We consider LiveATC-Test as low quality speech data set i.e., signal-to-noise (SNR) ratios goes from 5 to 15\,dB~\cite{zuluaga2020automatic}. Audio is sampled at 16\,kHz.

\subsection{Public databases}

\qquad \textbf{ATCO2-Test:} evaluation set released at Interspeech 2021 by~\cite{zuluagagomez21_interspeech, kocour21_interspeech}.
The dataset contains a mix of noisy and heavily English accented recordings from seven different airports located in Australia, Czech Republic, Slovakia, and Switzerland. The first version of the ATCO2-Test set contains 1.1\,hrs of speech, it is open-source and can be accessed for free in \url{https://www.atco2.org/data}. The full corpus is available for purchase through ELDA in \url{http://www.elra.info/en/catalogues/}. We only use the open-source version for reproducibility. The recordings of both corpus are mono-channel sampled at 16kHz and 16-bit PCM. 
This is the first study that evaluates E2E ASR for ATCO2-Test. The WERs listed in this paper could be adopted as baselines for future research.

\vspace{0.1cm}

\textbf{LDC-ATCC}: public ATC corpus gathered from three different airports.\footnote{\url{https://catalog.ldc.upenn.edu/LDC94S14A}}
LDC-ATCC corpus comprises recorded speech that aims to support research in robust ASR. The recordings contain several speakers and gathered over noisy channels. The dataset is formatted in NIST Sphere format, where full transcripts, start and end times of each transmission are provided. The audio files are sampled at 8\,kHz, 16-bit PCM~\cite{LDC_ATCC}. 
\vspace{0.1cm}

\textbf{UWB-ATCC}: free public ATC corpus containing recordings of communication between ATCOs and pilots. The speech is manually transcribed and labeled with speaker roles. The audio data is single channel sampled at 8\,kHz. This dataset can be downloaded for free in their website\footnote{\url{https://lindat.mff.cuni.cz/repository/xmlui/handle/11858/00-097C-0000-0001-CCA1-0}.}~\cite{UWB_ATCC}.
\vspace{0.1cm}

\textbf{ATCOSIM}: free public database for research on ATC communciations. It consists of 10\,hrs of speech data recorded during ATC real-time simulations using a close-talk headset microphone. The utterances are in English language and pronounced by ten non-native speakers. The database includes orthographic transcriptions and additional information about speakers and recording sessions. This dataset can be downloaded for free in their website{\footnote{\url{https://www.spsc.tugraz.at/databases-and-tools}}}~\cite{ATCOSIM}.

For all of our experiments, we up-sample all recordings to 16\,kHz. Additionally, there are not any official train/dev/test splits for LDC-ATCC, UWB-ATCC and ATCOSIM databases. Therefore, we split them following the proportions in Table~\ref{tab:datasets}. We also make sure that there is no speaker or utterance overlaps between each subset.

\begin{table}[t]
  \caption{Characteristics of public and private databases in the domain of air traffic control communications. We only list databases used in the experiments, see~\cite{zuluagagomez21_interspeech} for a more exhaustive list. $^\dagger$baseline performance of our state-of-the-art hybrid-based ASR model for ATC communications.}
  \vspace{0.1cm}
  \label{tab:datasets}
  \centering
  \begin{tabular}{ lccc }
    \toprule
    & \multicolumn{3}{c}{\cellcolor{Gray} \textbf{Characteristics}} \\
    \rule{0pt}{3ex} \textbf{Dataset} & \multicolumn{1}{c}{Train / Test} & \multicolumn{1}{c}{SNR [dB]} & \multicolumn{1}{c}{WER [\%]$^\dagger$} \\
    \midrule
    \rowcolor{Gray}
    \multicolumn{4}{c}{\textbf{\textit{Private databases}}} \\
    \midrule
    {\footnotesize\textbf{NATS}} & 18h / 0.9h & $\geq$20 & 7.7\\
    {\footnotesize\textbf{ISAVIA}} & 14h / 1h & 15-20 & 12.5 \\
    {\footnotesize\textbf{LiveATC-Test}} & - / 1.8h & 5-15 & 35.8 \\
    \midrule
    \rowcolor{Gray}
    \multicolumn{4}{c}{\textbf{\textit{Public databases}}} \\
    \midrule
    {\footnotesize\textbf{ATCO2-Test}} & - / 1.1h & 10-15 & 24.7 \\
    {\footnotesize\textbf{LDC-ATCC}} & 23h / 2.6h & 10-15 & - \\
    {\footnotesize\textbf{UWB-ATCC}} & 10.4h / 2.6h & $\geq$20 & - \\
    {\footnotesize\textbf{ATCOSIM}} & 8h / 2.4h & $\geq$20 & - \\
    \bottomrule
  \end{tabular}
\end{table}

\subsection{Automatic speech recognition}
\label{subsec:asr}

Our experimental setup is split into three parts, which aims to answer each of the questions raised in the Section~\ref{sec:intro}. Initially, we assess WERs of several E2E models when fine-tuned with ATC audio. We define two training datasets, i) 32\,hrs of annotated data from NATS and ISAVIA database and ii) 132\,hrs of ATC speech data from different projects (including all the training data from Table~\ref{tab:datasets}), and we redirect the reader to~\cite{zuluagagomez21_interspeech} for further details. For now on, we refer to these datasets as \textit{32\,hrs} and \textit{132\,hrs} `fine-tuning sets'. 
Later, we evaluate the low-resource scenario by fine-tuning E2E models with different amount of data, for this, we use NATS and ISAVIA as private databases, and LDC-ATCC and UWB-ATCC as public databases.
Finally, we evaluate the performance shift by fine-tuning E2E models with audio data from different genders. Here, we employ the free and open-source ATCOSIM database. Finally, we release training scripts to replicate the results of the last two experiments (only for the public databases).
\vspace{0.1cm}

\textbf{Baseline hybrid-based ASR}: all experiments are conducted with Kaldi toolkit~\cite{povey2011kaldi}. The baseline models are composed of six convolution layers and 15 factorized time-delay neural network ($\sim$31M trainable parameters). We follow the standard Kaldi's chain LF-MMI training recipe~\cite{povey2016purely}. The input features are high-resolution MFCCs with online cepstral mean normalization. The features are extended with i-vectors. We use 3-gram ARPA LM during decoding. The model is trained for 5 epochs on 132\,hrs of ATC speech (that includes NATS and ISAVIA). Further information and baseline performances can be found in our previous work~\cite{zuluagagomez21_interspeech,zuluagagomez20_interspeech,zuluaga2020automatic}. SOTA WERs are listed in the last column of Table~\ref{tab:datasets}.
\vspace{0.1cm}

\textbf{End-to-end ASR}: we report results on four configurations of Wav2Vec 2.0/XLS-R models. From now on, we refer to these models with the following tags: \mbox{i)~\textit{w2v2-B:}} BASE model (95M parameters, pre-trained on train-set 960\,hrs LibriSpeech~\cite{panayotov_librispeech_ICASSP2015}); ii)~\textit{w2v2-L:} LARGE-960h model (317M parameters pre-trained and then fine-tuned with LibrSpeech 960\,hrs train-set); iii)~\textit{w2v2-L-60K:} LARGE-960h-LV60K model (same as w2v2-L but uses LibriSpeech + 60k\,hrs from LibriVox project i.e., Libri-Light~\cite{kahn2020libri} during the pre-training phase); iv)~\textit{w2v2-XLS-R:} XLS-R model (300M parameters pre-trained on 436k\,hrs of publicly available data in 128 languages~\cite{babu2021xls}). We fetched all models' checkpoints from HuggingFace platform~\cite{wolf-etal-2020-transformers,lhoest2021datasets}. Later, we perform standard fine-tuning with ATC speech data.
\vspace{0.1cm}

\textbf{Hyperparameters end-to-end ASR}: all experiments use the same set of hyperparameters. The feature extractor is frozen throughout the fine-tuning phase. We fine-tune each model for 10\,k~steps, with a 500-step warm-up phase ($\sim$5\% of total updates). Learning rate is increased linearly until $\gamma=1\mathrm{e}{-4}$ during warm-up, then it linearly decays. We use CTC loss function~\cite{graves2006connectionist}. Dropout~\cite{srivastava2014dropout} is set to $dp=0.1$ for the attention and hidden layers. We use GELU activation function~\cite{hendrycks2016gaussian} and AdamW~\cite{loshchilov2018decoupled} optimizer ($\beta_1{=}0.9, \beta_2{=}0.999$, $\epsilon{=}1\mathrm{e}{-8}$). 
We fine-tune each model on an NVIDIA GeForce RTX 3090 with an effective batch size of 72 (batch size of 24, gradient accumulation of 3). 
All the models use a character-based lexicon, i.e., we concatenate the English alphabet with some symbols and the blank symbol (standard in CTC-based AM). In total, our vocabulary is composed of 32 characters, thus, we append a linear layer of this dimension on top of the pre-trained w2v2/XLS-R AMs to generate logits at each time step. We use greedy decoding after applying Softmax to obtain the most likely character at each time step.
\vspace{0.1cm}

\textbf{Language model (LM):} we concatenate all text transcripts and train 2/3/4-gram ARPA LMs. The LMs are integrated by shallow fusion with a Python based CTC decoder, \texttt{PyCTCDecode}. \footnote{\scriptsize{ \url{https://github.com/kensho-technologies/pyctcdecode}}} 4-gram LMs performed systematically better ($\sim$2\% relative WER reduction) compared to 2-gram LMs in all test sets. We report results only with 4-gram LM as in~\cite{baevski2020wav2vec}. We set $\alpha=0.5$ and $\beta=1.5$, which corresponds to the LM and length normalization weights. We set the beam size to 100. 
\vspace{0.2cm}

\textbf{Data augmentation:} we apply a data augmentation strategy similar to SpecAugment~\cite{park2019specaugment} is applied. We mask the input sequence with a probability $p=0.075$, and $M=12$ consecutive frames. These hyperparameters follow closely the original Wav2Vec 2.0 implementation~\cite{baevski2020wav2vec}.

\subsection{Incremental training}

\begin{figure}[t]
  \centering
  \includegraphics[width=0.99\linewidth]{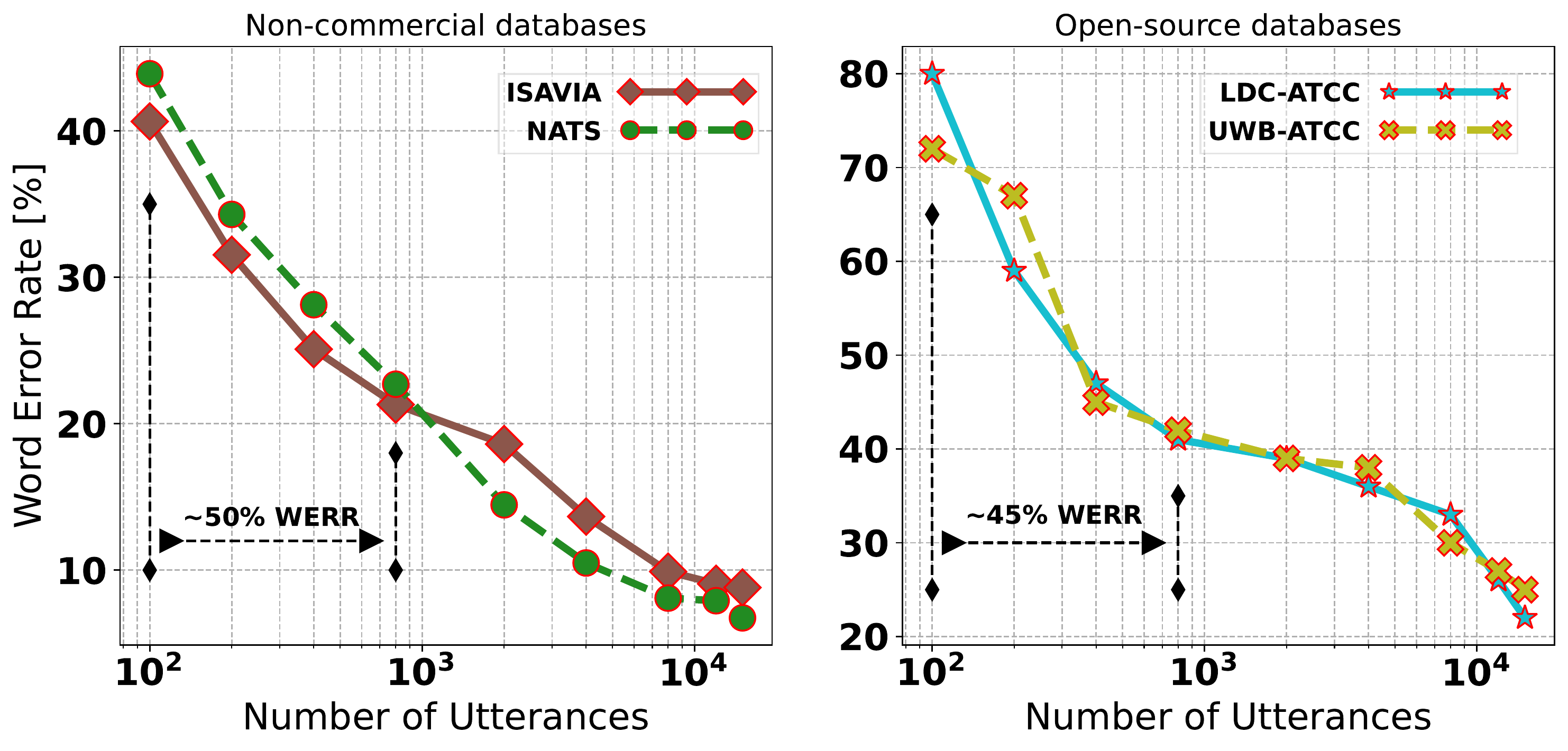}
  \caption{Word error rates (WER) in percentages (\%) for models fine-tuned with different amounts of data (x-axis). The left plot covers the results for private databases, while the right plot the public ones. Each data point corresponds to a train/test subset from the same dataset. 100, 1k and 10k utterances are roughly 5\,min (few-shot), 1\,h, and 10\,hrs, respectively. All the evaluations are reported with \textit{w2v2-L-60K} model and without explicit language model. We also list the Word Error Rate Reduction (WERR) by scaling up the fine-tuning set size from 100 to 800 samples.} 
  \label{fig:incremental_results}
  \vspace{-4mm}
\end{figure}

With the recent success of SSL pre-trained E2E models, it has become of particular interest to quantify how much data is actually needed to perform effectively on a downstream task. It is also important for low-resource tasks, such as ATC, where few tens of hours of labeled data are available for training or fine-tuning. In most ATC cases, data from one airport does not generalize well to other airports (for instance, see Table~\ref{tab:commercial-transfer}) due to a considerable AM domain-shift (accent, speaker rates and audio quality), as well as a LM domain-shift (dominance of different vocabulary). We analyze model performance versus different fine-tuning data sizes. We experimented with four \textit{few-shot learning} scenarios with less than one hour ($\sim$1k utterances) of fine-tuning data. We split the experiments in two. First, we fine-tuned nine models on private databases, either NATS or ISAVIA data, as depicted on the left plot of Figure~\ref{fig:incremental_results} (x-axis refers to number of utterances used during fine-tuning in log scale). Second, with the aim of open research, we performed the same approach on public databases, i.e., LDC-ATCC and UWB-ATCC. The results are on the right plot of Figure~\ref{fig:incremental_results}.

\subsection{Gender experiments}

We use the free and open-source ATCOSIM database to carry the gender experiments. We obtained the gender labels for each utterance from the original ATCOSIM gold annotations (check our public GitHub repository for more details). We split the train set into increasing sizes of 1h, 2h, 3h, 3.5h, and also by gender. We aim at both, analyzing the performance in WERs caused by fine-tuning an E2E with audio from different gender, and to measure the performance gain by scaling up the fine-tuning data.
We trained four models for each gender (using the same hyperparameters as the ones described in Section~\ref{subsec:asr} and with the same model: \mbox{\textit{w2v2-L-60k}}) and report the results in Table~\ref{tab:gender_bias}.

\section{Results and Discussion}
\label{sec:results}

\begin{table*}[th!]
    \caption{Word error rates (WER) in percentages (\%) on four ATC test sets. Each model is fine-tuned on NATS and ISAVIA data ($\sim$32\,hrs). WERs are reported with greedy decoding or beam search decoding with a 4-gram ARPA LM integrated by shallow fusion. Unlabeled data column: \textit{LS} stands for LibriSpeech 960\,hrs train-set~\cite{panayotov_librispeech_ICASSP2015}, \textit{LV} for LibriVox 60k\,hrs train-set~\cite{kahn2020libri} and \textit{ML} for 436k\,hrs of multilingual speech data~\cite{babu2021xls}. $^*$reports the baseline WER of Wav2Vec 2.0 (Table 1 from~\cite{baevski2020wav2vec}) and XLS-R (Table 11 from~\cite{babu2021xls}) models on LibriSpeech \texttt{test-other} set when only fine-tuned on 10\,hrs of labeled data (comparable to our setup). $^\dagger$best Kaldi hybrid-based model (see~\cite{zuluagagomez21_interspeech, zuluagagomez20_interspeech}) trained with the same amount of data as ${\dagger\dagger}$. $^{\dagger\dagger}$models fine-tuned with 132\,hrs of ATC speech data (instead of 32\,hrs) and twice the number of steps, i.e., 20k. Numbers in \textbf{bold} refer to top WERs for models fine-tuned with the 32\,hrs set and \underline{underline} with 132\,hrs set.}
    \vspace{0.1cm}
    \label{tab:results}
    \centering
    \begin{tabular}{p{2.8cm}cccccccccc} 
        \toprule
        & \textbf{Unlabeled} & \multicolumn{2}{c}{\textbf{NATS}} & \multicolumn{2}{c}{\textbf{ISAVIA}} & \multicolumn{2}{c}{\textbf{ATCO2-Test}} & \multicolumn{2}{c}{\textbf{LiveATC-Test}} & LS$^*$ \\
        \textbf{Model {\footnotesize(num. params.)}} & \textbf{data} & {\footnotesize{Greedy}} & +LM & {\footnotesize{Greedy}} & +LM & {\footnotesize{Greedy}} & +LM & {\footnotesize{Greedy}} & +LM & - \\
        \midrule
        \rowcolor{Gray} \textbf{Baseline {\footnotesize(31M)}} & & & & & & & & & & \\
        \quad Hybrid-based $^\dagger$ & - & - & 7.7 & - & 12.5 & - & 24.7 & - & 35.8 & -\\
        \midrule
        \rowcolor{Gray} \textbf{BASE {\footnotesize(95M)}} & & & & & & & & & & \\
        \quad w2v2-B & LS & 10.7 & 8.4 & 12.5 & 10.1 & 45.6 & 40.1 & 48.1 & 42.2 & 7.8 \\
        \midrule
        \rowcolor{Gray} \textbf{LARGE  {\footnotesize(371M)}} & & & & & & & & & &  \\
        \quad w2v2-L & LS & 9.3 & 7.6 & 11.7 & 9.5 & 44.9 & 40.0 & 47.5 & 41.4 & 6.1  \\
        \quad w2v2-L-60k & LS+LV & \textbf{6.8} & \textbf{5.4} & \textbf{8.8} & \textbf{7.3} & 34.6 & 31.2 & 39.8 & 34.5 & 4.9 \\
        \quad w2v2-L-60k+$^{\dagger\dagger}$ & LS+LV & 9.3 & \underline{7.4} & 11.2 & 9.1 & 23.3 & 21.2 & 31.1 & 27.2 & -\\
        \midrule
        \rowcolor{Gray} \textbf{XLS-R {\footnotesize(300M)}} & & & & & & & & & &\\
        \quad w2v2-XLS-R & ML & 8.4 & 6.5 & 10.5 & 8.2 & 39.1 & 33.8 & 42.9 & 36.7 & 15.4 \\
        \quad w2v2-XLS-R+$^{\dagger\dagger}$ & ML & \underline{9.0} & \underline{7.4} & \underline{10.4} & \underline{8.3} & \underline{\textbf{22.8}} & \underline{\textbf{19.8}} & \underline{\textbf{29.7}} & \underline{\textbf{24.9}} & -\\
        \bottomrule
    \end{tabular}
\end{table*}

We structure the discussion of the results by addressing concrete questions. Our main hypothesis is that E2E models trained by SSL learn a robust representation of speech~\cite{baevski2020wav2vec} and perform well on downstream tasks, i.e., ASR or multilingual ASR~\cite{babu2021xls}. 
\vspace{0.2cm}

\noindent \textbf{Breaking the paradigm, hybrid-based or E2E ASR?} Although hybrid-based ASR modeling has been the default for several years, a new wave of E2E architectures pre-trained by SSL for joint AM and LM is taking its place. We compare E2E models to our best hybrid-based ASR trained with the 132\,hrs fine-tuning set on Kaldi (\textbf{Baseline}, first row, Table~\ref{tab:results}). 
For E2E AMs we select two models. First,~\textit{w2v2-L-60k} to evaluate NATS and ISAVIA test sets, which is only fine-tuned on the 32\,hrs set, i.e., in-domain data. Second,~\mbox{\textit{w2v2-XLS-R+}} for ATCO2-Test and LiveATC-Test test sets, which is trained on 132\,hrs of ATC speech data~\cite{zuluagagomez21_interspeech, zuluagagomez20_interspeech}. The 132\,hrs set is a more diverse set, and it was also used to train the hybrid-based baseline model. We obtained 30 and 41\% relative word error rate reduction (WERR) on NATS and ISAVIA when using~\mbox{\textit{w2v2-L-60k}} instead of our hybrid-based ASR baseline.
The improvement is considerable, even though the baseline model is trained on four times more data than \textit{w2v2-L-60k} (see Table~\ref{tab:results}). Similarly, \mbox{\textit{w2v2-XLS-R+}} (last row: Table~\ref{tab:results}) surpasses the hybrid-based model on all four test sets, but more significantly on the two most challenging, ATCO2-Test and LiveATC-Test sets. In total, 19 and 30\% relative WERR on ATCO2-Test and LiveATC-Test were obtained, respectively (hybrid-based $\rightarrow$ \textit{w2v2-XLS-R+}).
\vspace{0.1cm}

However, it is worth mentioning that hybrid-based ASR is still considered the default in many industrial applications due to some advantages over E2E models. Two examples are, hybrid-based ASR does not require high-performance computing (e.g., GPUs) to perform real-time inference, while E2E models relies heavily on GPUs for speed. Further, hybrid-based ASR can be easily adapted to streaming scenarios with minimum degradation on WERs. Yet, E2E models still involve considerable architectural modifications to reach on-par WERs~\cite{moritz2019streaming,narayanan2021cascaded,lu2022endpoint}.
\vspace{0.1cm}

\noindent \textbf{Does additional partly-in-domain data increases ASR performance?} We answer this question by comparing models fine-tuned either on the 132\,hrs or 32\,hrs set. The former set is a mix of public and private databases, while the latter is only NATS + ISAVIA, thus private. Note that NATS and ISAVIA are clean in-domain ATC speech corpora, i.e., considered as in-domain on the 32\,hrs set and partly-in-domain otherwise (132\,hrs set). Differently, ATCO2-Test and LiveATC-Test can be considered noisy and partly out-of-domain sets, i.e., airport, acoustic, and LM mismatch. 

To address this question, we only focus on \textit{w2v2-L-60k} and \textit{w2v2-L-60k+} models fine-tuned on the 32\,hrs and 132\,hrs sets, respectively.\footnote{Note that the results are still comparable for the XLS-R AM, i.e., \mbox{\textit{w2v2-XLS-R}} versus \textit{w2v2-XLS-R+}.}. We analyze the WERs obtained by greedy decoding to focus only on joint acoustic and language ASR modeling (see Section~\ref{sec:related-work}). A degradation on WERs is observed for the in-domain test sets, NATS: 6.8\%~$\rightarrow$~9.3\% WER and ISAVIA: 8.8\%~$\rightarrow$~11.2\% WER. This is mainly to the addition of data that does not match NATS and ISAVIA. Contrary, there was considerable WERR on the partly out-of-domain sets, ATCO2-Test:~34.6\%~$\rightarrow$~23.3\% WER and LiveATC-Test~39.8\%~$\rightarrow$~31.1\% WER. NATS test set (ISAVIA: 1\% relative WERR) was impacted by the addition of partly-in-domain data, i.e., $\sim$7\% relative lower WERs. Nevertheless, challenging test sets improved dramatically, i.e., ATCO2-Test and LiveATC-Test 43\% and 33\% relative WERR.
\vspace{0.2cm}

\noindent \textbf{Do multilingual pre-trained E2E models help?} To answer this question we compare \textit{w2v2-L-60k+} and \textit{w2v2-XLS-R+} models, which use the same hyperparameters, fine-tuning setup and beam search decoding with LM. We obtain a relative WERR of 8.8\%, 6.6\% and 8.5\% on ISAVIA, ATCO2-Test and LiveATC-Test, respectively (no improvement on NATS). Significant improvement is seen on the most challenging test sets (SNR: 5-10\,dB) which contain accented English speech, i.e., ATCO2-Test and LiveATC-Test. 
Hence, multilingual pre-trained models bring a tiny, but noticeable boost in performance compared to single-language pre-trained E2E models. 
This observed behavior can be attributed to the fact that \mbox{\textit{w2v2-XLS-R}} have seen considerably more multilingual and accented audio data during the pre-training phase~\cite{babu2021xls} in comparison to \textit{w2v2-L-60k}~\cite{baevski2020wav2vec}.
\vspace{0.1cm}

The labeling process of ATC speech data can be decreased by a factor of two by only performing pre-labeling of audio with an in domain ASR system. Following this idea, we believe that is of special interest for the research community\footnote{Likewise, it is of special interest for the ATC community.} to understand and analyze how much audio data is needed to reach acceptable WERs. We validated this idea by performing experiments with different amounts of fine-tuning data, thus it is up to the interested party to define the `acceptable' WER threshold for the given application (e.g., deployment or pre-labeling).
\vspace{0.1cm}

\noindent \textbf{How much data do you need to fine-tune \mbox{Wav2Vec 2.0} and \mbox{XLS-R} models?} We also investigate the effect on WERs when different amounts of fine-tuning data are used during the fine-tuning phase. We divide the by public and private databases. The WERs on the private databases are in the left plot of Figure~\ref{fig:incremental_results}. All the experiments are based on the most robust E2E model from Table~\ref{tab:results} i.e., \textit{w2v2-L-60K}.\footnote{We select the \textit{best model} based on lowest WERs on out-of-domain test sets, i.e., ATCO2-Test and LiveATC-Test. See last row Table~\ref{tab:results}.} The WERs plot are obtained with greedy decoding and no LM or explicit textual information added. We fine-tune 18 models varying the training data set (either NATS or ISAVIA) and varying the amount of fine-tuning samples. We initially tested the few-shot learning scenario (`worse-case'), where only 100 labeled utterances ($\sim$5\,min) were used for fine-tuning, and achieved WERs of 40\% and 43.9\% for ISAVIA and NATS. Further, $\sim$50\% relative WERR is obtained by scaling up the fine-tuning data to 50 minutes (800 utterances). Specifically, NATS 43.9\% $\rightarrow$ 22.7\% WER and ISAVIA 40.6\% $\rightarrow$ 21.3\% WER. Lastly, if all available data ($\sim$14~hrs) is used, we reach an 8.8\% and 6.8\% WER for ISAVIA and NATS, respectively. This represents an $\sim$80\% relative WERR compared to the low-resource setup (100 utterances). With around 8\,hrs ($\sim$8000 utterances) \textit{w2v2-L-60K} beats the performance of our SOTA hybrid-based ASR (which uses four times more training data). We follow the same methodology to evaluate the public databases. We also train 9 models for each dataset, i.e., LDC-ATCC and UWB-ATCC. We list the WERs on the right plot of Figure~\ref{fig:incremental_results} for both test sets. Here, we note similar behaviors, thus we reach similar conclusions. First, scaling-up the fine-tuning data from 5 to 50 minutes brought $\sim$45\% relative WERR for both, LDC-ATCC and UWB-ATCC test sets (similar trend in private databases, NATS and ISAVIA). Not surprisingly, further gains in WERs are achieved if we increase the fine-tuning data up to 11\,hrs. Previous research has not explored E2E modeling\footnote{Training scripts to replicate the right plot of Figure~\ref{fig:incremental_results} are public in our GitHub repository.} in the area of ATC, thus, these WERs can be adopted as baselines.

\vspace{0.1cm}

\begin{table}[t]
  \caption{Word error rates (WER) in percentages (\%) on different test sets. Models are fine-tuned only on public databases and fixed to 11\,hrs of audio data. All systems are \mbox{\textit{w2v2-L-60k}} and WERs are obtained with greedy decoding and no LM. $^{\dagger}$test set split by gender (Male/Female).}
  \vspace{0.1cm}
  \label{tab:commercial-transfer}
  \centering
  \begin{tabular}{ lccccc }
    \toprule
     &  \multicolumn{4}{c}{\cellcolor{Gray} \textbf{Test set}} \\
    \textbf{Train set} & {\footnotesize LDC} & {\footnotesize UWB} & ATCO2 & {\footnotesize ATCOSIM (M/F)$^{\dagger}$} \\
    \midrule
    {\footnotesize LDC-ATCC} & \textbf{25.0} & 64.1 & 58.7 & 41.1 / 35.7 \\
    {\footnotesize UWB-ATCC} & 54.6 & \textbf{21.9} & \textbf{47.9} & \textbf{32.5} / \textbf{24.6} \\
    \bottomrule
  \end{tabular}
\end{table}

\noindent \textbf{Transferability between ATC corpora:} we have stated before that E2E models fine-tuned on a specific ATC corpus might not transfer well to different ATC corpora.\footnote{This assumption also applies to hybrid-based ASR models.} To test this premise, we train models with different public databases and test them on 4 test sets. We fixed the model (\mbox{\textit{w2v2-L-60k}}), training data size to 11\,hrs, and same hyperparameters. From Table~\ref{tab:commercial-transfer}, we can conclude that UWB-ATCC corpus transfers better to different databases, for instance LDC-ATCC. In this case, if we fine-tune \mbox{\textit{w2v2-L-60k}} with UWB-ATCC set and test it on LDC-ATCC the performance is 54\% WER, whereas inversely the performance is 64\%, i.e., $\sim$10\% absolute WERR. Similarly, the model trained on UWB-ATCC fits better ATCO2 test by a large margin compared to LDC-ATCC, i.e., 10\% absolute lower WER. 

\vspace{0.1cm}

\noindent \textbf{Gender bias on air traffic control communications:} finally, it is becoming a trend to go beyond simply analyzing the performance of E2E models on the selected downstream task, e.g., ASR. For instance, examining the performance shift of a given ASR model when fine-tuned on audio from different genders~\cite{meng2022don,zanonboito22_interspeech}. We analyze this bias on ATCOSIM dataset, which provide the gender labels for each utterance. The results are listed in Table~\ref{tab:gender_bias}. It is evident that the experiments with female voice performed systematically better in all training scenarios (1h to 3.5h fine-tuning set). We also wanted to rule out the possibility that the speech rate was the main cause of this behavior. In average, each female recording has a speech rate of 3.4 words per second (WPS) while male has an average of 2.9 WPS. In order to determine whether female recordings are of better quality than the male ones, or whether the E2E model have some bias acquired during the pre-training phase, we calculated the WERR when fine-tuning the model between 1h to 3.5h of audio. Following Table~\ref{tab:gender_bias} we can see that in the female experiments the reduction on WERs is higher than on the male side by around 8\% absolute when scaling from 1h to 3.5h. 

We believe that these E2E models (e.g., Wav2Vec 2.0) might carry little but noticeable gender bias. This could be one reason why the experiments with female recordings performed better. For instance, previous work have concluded that gender unbalance might affect E2E models during the pre-training phase. However, this bias can be mitigated by adding a small amount of data from the opposite gender~\cite{meng2022don}. In conclusion, it is still prudent to perform more thorough experiments before reaching hard judgments in this regard, or at least, in ATC communications.


\section{Conclusion}

This paper evaluated the robustness of pre-trained Wav2Vec 2.0 and XLS-R models on downstream ASR for ATC on different corpora. 
Our experiments show large recognition improvements of Wav2Vec 2.0 and XLS-R compared to \textit{hybrid-based} ASR baselines. Quantitatively, between 20\% and 40\% relative WERR was obtained on ISAVIA and NATS test sets, but also from challenging multi-accent databases i.e., ATCO2-Test and LiveATC-Test. Furthermore, we demonstrated that pre-trained Wav2Vec 2.0 models allow rapid fine-tuning with small quantities of adaptation data. For instance, $\sim$5\,min of speech allows fine-tuning a model that yields WERs of 40\% and 43.9\% for ISVAIA and NATS, respectively. Moreover, we showed that at least 4\,hrs of in-domain data already provide acceptable WERs of $\sim$10\% for ISAVIA and NATS recordings and by using two times more data (i.e., 8\,hrs) performance surpasses hybrid-based ASR baselines. 
We also performed the same analysis for public databases. Similar WERs ($\sim$40\%) are achieved with 50\,min of data. 
Finally, the gender bias in one ATC corpus is also covered by training gender dependent ASR models. We found that large-scale speech models perform systematically better on female recordings and also more gains in WERs are achieved when scaling up the fine-tuning data, in comparison to male recordings. 
The strength of this research is that, this is the first research aiming at analyzing the performance of these large-scale speech models on ATC. In addition, this is the first work in the ATC area that is publicly releasing a GitHub repository to replicate experiments.

\begin{table}[t]
  \caption{Word error rates (WER) in percentages (\%) on ATCOSIM dataset with different fine-tuning set sizes. WERs are reported on 0.7\,hrs of speech (only from the same gender) sampled from the original test set, i.e., we fine-tune and evaluate within the same gender. All models are trained with \mbox{\textit{w2v2-L-60k}} and decoding is done without LM. $^\dagger$list the word error rate reduction (WERR) achieved by scaling from 1h to 3.5h of speech during the fine-tuning stage.}
  \vspace{0.1cm}
  \label{tab:gender_bias}
  \centering
  \begin{tabular}{ lcccccc }
    \toprule
     &  \multicolumn{4}{c}{\cellcolor{Gray} \textbf{Dataset size}} & \textbf{WERR}$^\dagger$ \\
    \textbf{Gender} & 1h & 2h & 3h & 3.5h & (1h $\rightarrow$3.5h)\\
    \midrule
    Male & 36.70 & 31.42 & 29.20 & \textbf{28.72} & 21.74\% \\
    Female & 17.62 & 13.91 & 13.46 & \textbf{12.37} & 29.79 \% \\    
    \bottomrule
  \end{tabular}
\end{table}

\section{Acknowledgments}

The work was supported by SESAR Joint Undertaking under Grant Agreement No. 884287 - HAAWAII (\textbf{H}ighly \textbf{A}utomated \textbf{A}ir traffic controller \textbf{W}orkstations with \textbf{A}rtificial \textbf{I}ntelligence \textbf{I}ntegration). The work was also partially supported by CleanSky Joint Undertaking under Grant Agreement No. 864702—ATCO2 (Automatic collection and processing of voice data from air-traffic communications).

\bibliographystyle{IEEEbib}
\bibliography{mybib}

\end{document}